\documentclass[epj,twocolumn]{webofc}
\usepackage[varg]{txfonts}   
\woctitle{CGS15} 
\begin{document}
\title{Nucleosynthesis simulations for the production of the p-nuclei $^{\text{92}}$Mo and $^{\text{94}}$Mo in a Supernova type II model}

\author{Kathrin G\"obel\inst{1,4}\fnsep\thanks{\email{goebel@physik.uni-frankfurt.de}} \and 
Jan Glorius\inst{1,2} \and
Alexander Koloczek\inst{1,4} \and
Marco Pignatari\inst{3,4} \and
Ren\'e Reifarth\inst{1,4} \and
Ren\'e Schach\inst{1,4} \and
Kerstin Sonnabend\inst{1} }

\institute{Goethe University Frankfurt am Main, Germany \and
GSI Helmholtzzentrum für Schwerionenforschung, Darmstadt, Germany \and
University Basel, Switzerland \and
NuGrid collaboration, \url{http://www.nugridstars.org}}

\abstract{We present a nucleosynthesis sensitivity study for the $\gamma$-process in a Supernova type II model within the NuGrid research platform. The simulations aimed at identifying the relevant local production and destruction rates for the p-nuclei of molybdenum and at determining the sensitivity of the final abundances to these rates. We show that local destruction rates strongly determine the abundance of $^{92}$Mo and $^{94}$Mo, and quantify the impact.}
\maketitle
\section{Introduction}
\label{intro}
The p-nuclei between $^{74}$Se and $^{196}$Hg can be produced under explosive conditions in a sequence of photodissociation reactions and subsequent $\beta$-decays starting at s- and r-process seeds~\cite{1992AARvLambert}. Most of the p-nuclei are about two orders of magnitude less abundant than other stable isotopes belonging to the same element. Relevant exceptions are the isotopes $^{\text{92,94}}$Mo and $^{\text{96,98}}$Ru~\cite{2003arnould}. Their production in stars is a puzzle for nuclear astrophysics since present models underproduce these p-nuclei by orders of magnitude. According to recent stellar model calculations, $^{\text{94}}$Mo is mainly synthesized via the ($\gamma$,n) photodisintegration chain starting from the more neutron-rich and stable molybdenum isotopes~\cite{Rauscher2006PhRvCBranchings}. Some of the light p-nuclei, like the neutron magic isotope $^{\text{92}}$Mo, may also be synthesized by proton capture reactions~\cite{2011ApJTravaglio}.

We investigate the nucleosynthesis of $^{\text{92}}$Mo and $^{\text{94}}$Mo in a Supernova type II model. We present results on the sensitivity of rates of direct production or destruction reactions of the two Mo isotopes.

\section{PPN simulations using a Supernova type II model}
\label{secPPNSimSNII}

\begin{figure}[h]
\centering
\includegraphics[trim=0 10 30 40, clip=true, width=1.0\linewidth]{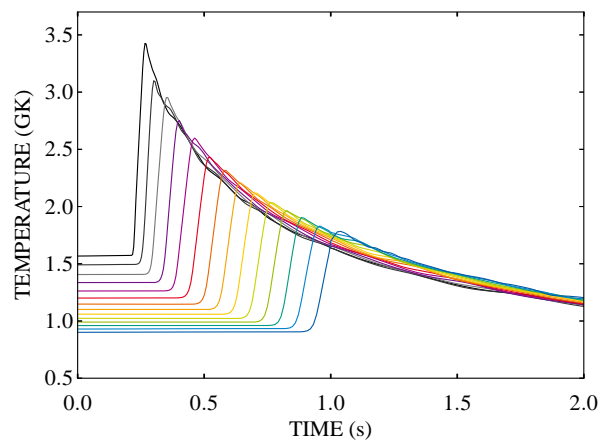}
\caption{Temperature profiles of the classic Supernova type II trajectories~\cite{2006ApJRapp}. Left to right: profiles from the innermost mass layer to the outermost.}
\label{fig:HashTemp}
\end{figure}

For the post-processing nucleosynthesis (PPN) simulations we use a stellar progenitor with an initial mass of 25 $M{}_{\odot}$ and metallicity \mbox{$Z$ = 0.02}~\cite{NuGridI}. The p-process nucleosynthesis is calculated by using classic Supernova type II trajectories~\cite{2006ApJRapp}. In the model, a shock front passes through the Ne/O burning zone, which is subdivided into 14 mass layers. Figure~\ref{fig:HashTemp} shows the temperature profiles, which describe the astrophysical environment of each layer as a function of time. The innermost mass shell is the hottest and densest environment with a peak temperature of \mbox{3.45 GK}. The maximum value drops to \mbox{1.79 GK} for the outermost mass layer. The shock front reaches the mass layers successively. The temperature rapidly increases to the maximum value and afterwards drops slowly.

The complete nucleosynthesis is calculated using the post-processing code PPN within the NuGrid research framework~\cite{HerwigNICX2008,NuGridNPN2012}. NuGrid offers a software framework for nucleosynthesis simulations in relevant astrophysical environments. A large network of over 5,000 isotopes and more than 60,000 reactions is used. Most of the nuclear reaction rates adopted are taken from the JINA Reaclib Database V1.1~\cite{JINACyburt2010}.

\section{Nucleosynthesis fluxes}
\label{secNuclFluxes}

\begin{figure}
\centering
\includegraphics[trim=0 370 0 250, clip=true,width=1.0\linewidth]{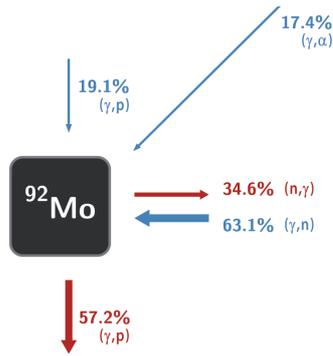}
\caption{Time-integrated nucleosynthesis fluxes producing and destroying the isotope $^{\text{92}}$Mo. The sum of the production fluxes is normalized to 100\%, and the destruction fluxes are scaled with the same factor. Fluxes smaller than 1\% are not shown.}
\label{fig:IntegrFluxes92Mo}
\end{figure}

\begin{figure}
\centering
\includegraphics[trim=0 150 0 250, clip=true,width=1.0\linewidth]{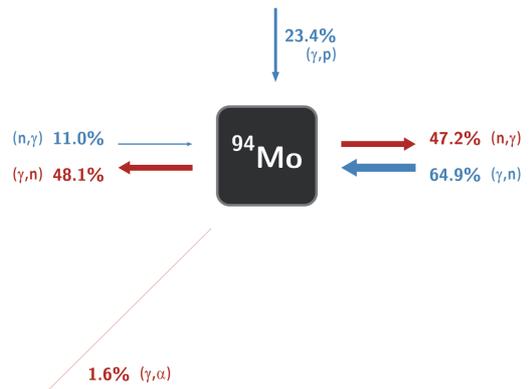}
\caption{Same as Figure~\ref{fig:IntegrFluxes92Mo}, for $^{\text{94}}$Mo.}
\label{fig:IntegrFluxes94Mo}
\end{figure}

The relative time-integrated nucleosynthesis fluxes of all mass layers for the isotopes $^{\text{92}}$Mo and $^{\text{94}}$Mo are shown in Figures~\ref{fig:IntegrFluxes92Mo} and~\ref{fig:IntegrFluxes94Mo}. For each isotope, the sum of the production fluxes is normalized to 100\%, and the destruction fluxes are scaled with the same factor. Fluxes smaller than 1\% are not shown. If the sum of destruction fluxes is less than 100\%, there is a net yield in the simulation. Otherwise the isotope gets depleted. The sum of the production fluxes is larger than the sum of the destruction fluxes for both isotopes. The isotope $^{\text{92}}$Mo has the largest net yield with 100\% production and 93\% destruction. 

The isotope $^{\text{92}}$Mo is mainly produced by photodisintegration of $^{\text{93}}$Mo. The reaction $^{\text{92}}$Mo($\gamma$,p) is the main destruction path followed by $^{\text{92}}$Mo(n,$\gamma$) reactions. The reaction $^{\text{95}}$Mo($\gamma$,n) strongly contributes to the production fluxes of $^{\text{94}}$Mo. The isotope $^{\text{94}}$Mo is equally destroyed by ($\gamma$,n) and (n,$\gamma$) reactions. 

The reactions accounting for a significant fraction of the nucleosynthesis fluxes are investigated in the rate variation study in section~\ref{secRateVariationStudySNII}.


\section{Abundances of $^{\text{92}}$Mo and $^{\text{94}}$Mo depending on local production and destruction rates}
\label{secRateVariationStudySNII}

The PPN simulations using a Supernova type II model aim at identifying the relevant reactions for the production and destruction of the Mo p-nuclei, and at quantifying the effect of rate changes on the final abundances. Previous nuclear sensitivity studies varied the reaction rates for all reactions of a certain type in the network, as well as for all reactions involving the p-nuclei simultaneously~\cite{2006ApJRapp,Rauscher2006PhRvCBranchings}. In this work, we explore the impact of single rates directly producing or destroying $^{\text{92}}$Mo and $^{\text{94}}$Mo. 

The effect of single rate variations on the final abundances of the p-nuclei $^{92}$Mo and $^{94}$Mo was determined by multiple simulations of the same scenario using different reaction rates. The rates were multiplied/divided by factors 2 and 5 for reactions involving a neutron (($\gamma$,n) and (n,$\gamma$)) or a proton (($\gamma$,p) and (p,$\gamma$)), and by factors 5 and 10 for reactions involving an $\alpha$-particle (($\gamma$,$\alpha$) and ($\alpha$,$\gamma$)). The factors were chosen according to the commonly stated uncertainties of the reaction rates (\cite{Rauscher2006PhRvCBranchings} and references therein). In the PPN simulation, the rate taken from the library is calculated for the current temperature in the astrophysical environment and then multiplied by the factor set by the user.

This work shows the dependencies for the rates where the maximum abundance ratio is larger than 5\% within the applied rate variations. 

\subsection{The case of $^{\text{92}}$Mo}\label{sec:92MoResults}

\begin{figure}
\centering
\includegraphics[trim=0 40 0 30, clip=true, width=1.0\linewidth]{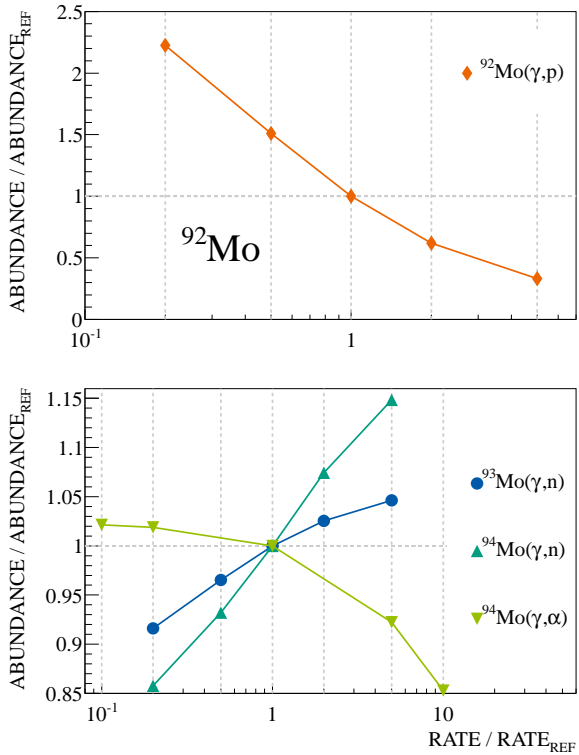}
\caption{Impact on the abundance of $^{\text{92}}$Mo depending on the $\gamma$-induced production and destruction rates. The results for the reaction
rate of $^{\text{92}}$Mo($\gamma$,p) (top) are shown in a single plot due to its large impact on the abundance compared to the other reaction rates.}
\label{fig:Mo92aburatio}
\end{figure}

Figure~\ref{fig:Mo92aburatio} shows the final abundance of $^{\text{92}}$Mo as a function of the factor applied to one production or destruction rate. The abundance of $^{\text{92}}$Mo strongly depends on the $^{\text{92}}$Mo($\gamma$,p) reaction rate, which is the main destruction path. The abundance of $^{\text{92}}$Mo drops with increasing rate and vice versa (Figure~\ref{fig:Mo92aburatio}, top panel). Neutron dissociation reactions are the main production path of $^{\text{92}}$Mo (Figure~\ref{fig:Mo92aburatio}, bottom panel). The $^{\text{94}}$Mo($\gamma$,n) reaction shows a strong impact on the abundance of $^{\text{92}}$Mo. Variations of the reaction rate of $^{\text{93}}$Mo($\gamma$,n) hardly affect the $^{\text{92}}$Mo abundance since the unstable isotope $^{\text{93}}$Mo has to be produced first. 

The $^{\text{92}}$Mo abundance drops if the reaction rate of $^{\text{94}}$Mo($\gamma$,$\alpha$) increases. The ratio of the reaction rates of $^{\text{94}}$Mo($\gamma$,$\alpha$) and $^{\text{94}}$Mo($\gamma$,n) is larger at high temperatures, hence, the nucleosynthesis path is altered producing less $^{\text{92}}$Mo via ($\gamma$,n) reactions. 

\subsection{The case of $^{\text{94}}$Mo}\label{sec:94MoResults}

Photodisintegration reactions starting from heavier Mo isotopes produce $^{\text{94}}$Mo. The destruction reaction $^{\text{94}}$Mo($\gamma$,n) mainly determines the final abundance of $^{\text{94}}$Mo. In the simulation, a small fraction of the nucleosynthesis fluxes proceeds via $^{\text{94}}$Mo($\gamma$,$\alpha$). An increase of the $^{\text{94}}$Mo($\gamma$,$\alpha$) rate reduces the abundance of $^{94}$Mo significantly. If the $^{\text{93}}$Mo($\gamma$,n) rate is increased, more $^{\text{93}}$Mo is destroyed and not available for the production of $^{\text{94}}$Mo via $^{\text{93}}$Mo(n,$\gamma$).  

\begin{figure}
\centering
\includegraphics[trim=0 0 0 20, clip=true, width=1.0\linewidth]{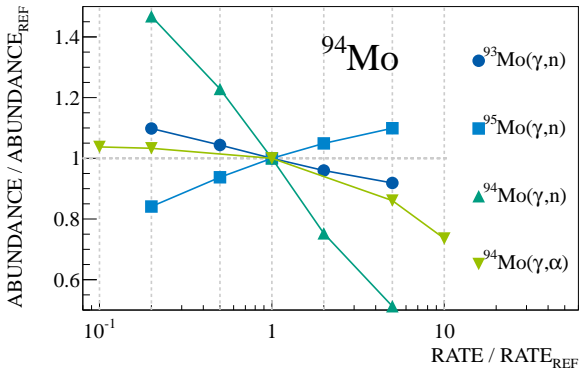}
\caption{Impact on the abundance of $^{\text{94}}$Mo depending on the $\gamma$-induced production and destruction rates.}
\label{fig:Mo94aburatio}
\end{figure}

\section{Summary and outlook}\label{secConclusions}

The impact of variations of local production and destruction rates on the p-nuclei $^{\text{92}}$Mo and $^{\text{94}}$Mo was investigated. The abundance of $^{\text{92}}$Mo is mainly determined by the $^{\text{92}}$Mo($\gamma$,p) reaction rate, the $^{\text{94}}$Mo abundance by the $^{\text{94}}$Mo($\gamma$,n) and $^{\text{94}}$Mo($\gamma$,$\alpha$) reaction rates. Most rates taken from JINA Reaclib Database V1.1 are calculated within the Hauser-Feshbach model~\cite{JINAonlineV11}. New evaluated data can be compared to these rates and new final abundances of the Mo p-nuclei can be obtained from the results presented here.  

In the $\gamma$-process, reactions on heavy nuclei ($Z$ > 42) may determine the reaction flow to the light p-nuclei and, hence, the abundance of $^{\text{92}}$Mo and $^{\text{94}}$Mo. Future studies foresee a global sensitivity study of reaction rates in different p-nuclei production scenarios including Supernovae type Ia. 

\begin{acknowledgement}
This project was supported by the Helmholtz International Center for FAIR, DFG (SO907/2-1) and HGS-HIRe. NuGrid acknowledges significant support from NSF grants PHY 02-16783 and PHY 09-22648 (JINA), EU MIRG-CT-2006-046520 and from the SNF (Switzerland).
\end{acknowledgement}

\bibliography{bibCGS15_KathrinGoebel}

\end{document}